\documentclass[10pt,onecolumn,letter]{IEEEtran}

\usepackage{amsmath,amssymb}
\usepackage{changepage}
\usepackage{textcomp,marvosym}
\usepackage[noadjust]{cite}
\usepackage{nameref,hyperref}
\usepackage[table]{xcolor}
\usepackage{graphicx}
\usepackage{array}
\usepackage{multicol}
\usepackage{color}
\usepackage{tabularx}
\usepackage[normalem]{ulem}
\definecolor{darkgreen}{rgb}{0,0.5,0}

\newcolumntype{+}{!{\vrule width 2pt}}

\newlength\savedwidth

\newcommand\thickhline{\noalign{\global\savedwidth\arrayrulewidth\global\arrayrulewidth 2pt}%
	\hline
	\noalign{\global\arrayrulewidth\savedwidth}}

\usepackage[aboveskip=1pt,labelfont=bf,labelsep=period,justification=raggedright,singlelinecheck=off]{caption}

\makeatletter
\renewcommand{\@biblabel}[1]{\quad#1.}
\makeatother

\date{}

\begin{document}
	
		\title{Accounting for the complex hierarchical topology of EEG phase-based functional connectivity in network binarisation} 
		
		\author{
			Keith Smith$^{1,2,*}$, Daniel Ab\'asolo$^{3}$, \& Javier Escudero$^{1}$			\thanks{$^{1}$Institute for Digital Communications, School of Engineering, University of Edinburgh, West Mains Rd, Edinburgh, EH9 3FB, UK}
			\thanks{$^{2}$Alzheimer Scotland Dementia Research Centre, Psychology Department, University of Edinburgh, 7 George Square, Edinburgh, EH8 9JZ, UK}
			\thanks{$^{3}$Centre for Biomedical Engineering, Department of Mechanical Engineering Sciences, Faculty of Engineering and Physical Sciences, University of Surrey, Guildford, United Kingdom}
			\thanks{$^{*}$PhD funded by EPSRC, e-mail: k.smith@ed.ac.uk}%
		}
		
	\maketitle
	\section*{Abstract}
	Research into binary network analysis of brain function faces a methodological challenge in selecting an appropriate threshold to binarise edge weights. For EEG phase-based functional connectivity, we test the hypothesis that such binarisation should take into account the complex hierarchical structure found in functional connectivity. We explore the density range suitable for such structure and provide a comparison of state-of-the-art binarisation techniques, the recently proposed Cluster-Span Threshold (CST), minimum spanning trees, efficiency-cost optimisation and union of shortest path graphs, with arbitrary proportional thresholds and weighted networks. We test these techniques on weighted complex hierarchy models by contrasting model realisations with small parametric differences. We also test the robustness of these techniques to random and targeted topological attacks.We find that the CST performs consistenty well in state-of-the-art modelling of EEG network topology, robustness to topological network attacks, and in three real datasets, agreeing with our hypothesis of hierarchical complexity. This provides interesting new evidence into the relevance of considering a large number of edges in EEG functional connectivity research to provide informational density in the topology.
	
	
	\section*{Introduction}
	Functional connectivity assesses the interdependent relationships between time-series recorded at spatially separated brain regions. Estimating and analysing functional connectivity using network science is an established methodology for extracting functional information from brain recordings taken using various platforms, most prominently the Electroencephalogram (EEG) and the Magnetoencephalogram (MEG) for high temporal resolution and functional Magnetic Resonance Imaging (fMRI) for high spatial resolution \cite{Bull2009}. Networks are proficient in their ability to capture the interdependent activity which underlies brain function \cite{Fall2014}, enabling powerful methods for classification of clinical patients \cite{Stam2014b} including in Alzheimer's disease (AD) \cite{Tijm2013} and Schizophrenia \cite{Calh2009}. Such analysis generally falls under the nomenclature of brain networks \cite{Bull2009}, which also includes studies of the brain's physical connections referred to as structural connectivity.
	
	Functional connectivity defined between all possible pairs of brain regions, whether sensors or cortical sources from the EEG or the MEG or partitioning of spatial regions in fMRI, present the researcher with a full adjacency matrix whose entries are only distinguished by relativity of weights. Since the most popular network science techniques are based on binary networks, initial efforts on the analysis of brain network topologies were implemented via the binarisation of the weights using some arbitrarily chosen thresholds with some good success \cite{Stam2007,Bull2009,Acha2007}. Still, studying the original weighted networks holds appeal in that it is more direct and has advantages of maintaining the information of relative strengths of connections \cite{Rubi2011}. However, these computed weights are known to vary due to any number of different pre-processing choices or connectivity analyses implemented, thus complicating comparisons and obfuscating results \cite{vanW2010,Stam2014b}. Furthermore, since the weights of dependency measures generally follow a non-scaling distribution between 0 and 1, many interesting topological considerations in binary networks, such as concerning paths, become redundant in light of the fact that the shortest weighted path between any two nodes is likely to be just the weighted edge connecting them \cite{Smit2016a}. Therefore, binarising the weights remains a more widely used approach which can explain the main topology of the underlying activity while alleviating methodological biasing and topological redundancy of weights.
	
	Selecting a method to binarise the network is thus seen as a major step in network construction in which the researcher is presented with a large degree of subjective choice \cite{vanW2010,Fall2014,Papo2014,vanD2015}. Because of this, recent research emphasises the importance of solutions to the thresholding or binarisation problem in functional connectivity \cite{Gin2011, Sch2011, Tew2014, Gar2015, Mei2015, vanD2015, Jal2016, Heuv2017, Fall2017}. While some find sparsity desirable based on the physiological hypothesis that function should be regarded as emerging through physically connected regions explained by a low wiring cost \cite{Tew2014,Acha2007,Fall2017}, others propose that less strong connections do not necessarily mean lower importance \cite{Schw2011,Sant2014} and, indeed, in real data analysis higher densities have been found to be as or more relevant \cite{Jal2016,Stam2007,Smit2015b}. In this study we focus on a particular case of phase-based connectivity obtained from EEG signals and hypothesise that the recently found complex hierarchical topology in this modality \cite{Smit2016b} contributes important information found only in higher densities.
	
\begin{figure}[!h]
		\centering
		\includegraphics[trim = 0 100 0 120,clip,scale=0.46]{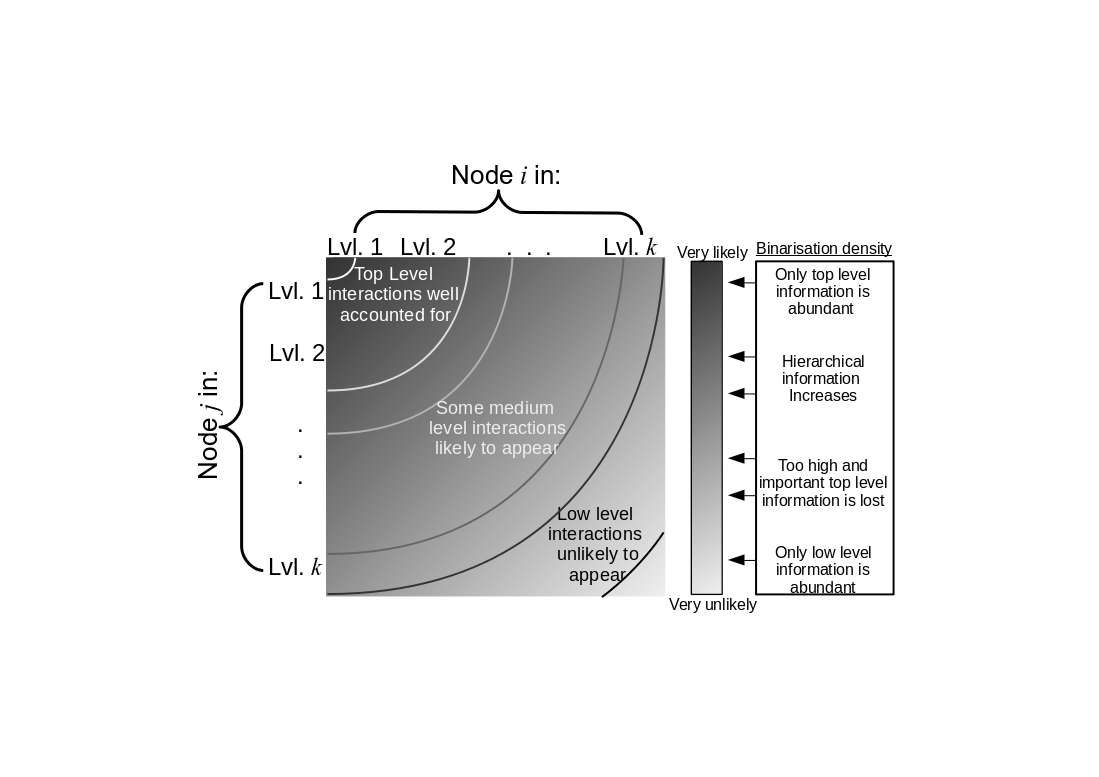}
		\caption{Illustration of the likelihood of an edge appearing between nodes $i$ and $j$, in hierarchy levels denoted by the $x$ and $y$ axes, in the binarised form of a weighted hierarchical network. Left, the effects of increasing binarised network density (strongest weights kept) on the hierarchical information of the network where black indicates 0\% density and white indicates 100\% density.}
		\label{edgelikelihood}
	\end{figure}

	In the case of EEG recorded at the scalp, one measures the electromagnetic activity of the brain after its propogation through the layers of bone and tissue of the head and is thus susceptible to noise and volume conduction effects. Nonetheless, its combination of practicality, portability and high temporal resolution make it a very appealing methodology. It has recently been uncovered that the hierarchical structure of EEG functional connectivity is a key aspect of its informational complexity \cite{Smit2016b}. Indeed, this article also showed that functional connectivity is characterised by high degree variance which is indicative of the large range of the general strength of network nodes. That is, one can expect that certain nodes have generally large adjacent weights, while others may have generally small adjacent weights. For a given node, the relativity of weight magnitudes of edges adjacent to the other nodes in the network is thus important to keep track of throughout the network and not just in the largest connections and nodes in the highest hierarchy levels as would be promoted in sparse densities, see Fig \ref{edgelikelihood}. Thus we propose that any useful binarisation technique for EEG phase-based functional connectivity should necessarily be able to account for the density of information inherent in a broad complex hierarchical structure. It should follow that sparsity is not necessarily a desirable feature of such brain functional networks and also that statistical thresholding on a case by case basis of the connectivity computations does not necessarily translate to a topological advantage in the resulting networks.
	
	To provide rigorous simulation results for binarisation techniques, we implement the Weighted Complex Hierarchy (WCH) model \cite{Smit2016b}. To our knowledge, this is the only generative weighted model of EEG functional connectivity to date which approximates the topological characteristics of EEG functional connectivity over the full range of densities. Moreover, the parameters of the model provide a fine-tuning of hierarchical topology which allows us to create a ground truth of subtly different topologies. Here we exploit this to assess the ability of binarisation methods to correctly identify topological differences between networks. This is intentionally done to echo the setup of a neuroscientific study and make the simulations as relevant as possible to the research community. This follows since, for the lack of a ground truth, studies in network neuroscience are generally based on contrasting conditions such as in cognitive tasks or contrasting populations where, for example, network features of patients are compared against those of healthy, age-matched controls \cite{Tijm2013}.
	
	Using simulations, we seek to clarify how network size and density range may effect the ability to discern small topological differences in network topology. In analysis, we compare state-of-the-art non-arbitrary binarisation techniques with a number of arbitrary percentage thresholds. Here we provide an extensive and full comparison of state-of-the-art non-arbitrary binarisation techniques- the Minimum Spanning Tree (MST) \cite{Tew2014,Stam2014}, Union of Shortest Paths (USP) \cite{Mei2015}, the Cluster-Span Threshold (CST) \cite{Smit2015a, Smit2016a}, Efficiency Cost Optimisation (ECO) \cite{Fall2017}. We also compare with weighted networks to directly compare binary and weighted approaches, as well as a number of arbitrary density thresholds for distinguishing differences. The MST and ECO create very sparse networks while the CST is notable in its production of consistently medium density networks (30-50\%) and the density of the USP depends on the distribution of edge weights \cite{Smit2016a}.
	
	We further analyse these techniques when the simulations are subject to random and targeted topological attack. We regard these as random and targeted attacks \cite{Tan2005} which preserve the network size. This is desirable given that many metric values are dependent on network size \cite{Dorm2009}. By randomising a percentage of weights in populations of subtly different complex hierarchical networks in parallel we can test how well the binarisation techniques can still uncover the differences between these populations, under varying sizes of `attack'. We implement these analyses to test the binarisation techniques' robustness in representing true network characteristics in the face of noise and/or outliers in the estimation of coupling between brain time series.
	
	We go on to apply our non-arbitary binarisation techniques to three real EEG datasets. We compare our thresholds on distinguishing the well known alpha activity existing between eyes open vs eyes closed resting state conditions in healthy volunteers with a 129 channel EEG \cite{NBTdata}. We then compare these techniques for distinguishing visual short-term memory binding tasks in healthy young volunteers with a 30 channel EEG \cite{Smit2015a}. Finally, we compare our techniques in distinguishing between AD patients and healthy control in a 16 channel EEG set-up \cite{EscData}. The varying sizes of these networks provides evidence for the translatability of the methods to different network sizes in the applied setting. The scripts, functions and data sets used in this study are available at the University of Edinburgh's data depository: http://datashare.is.ed.ac.uk/handle/10283/2783. 
	
	\section*{Methods}
	This section details the network simulations (A), binarisation techniques (B), network metrics (C), statistical tests (D) and real datasets (E) used in this study. Let $\mathbf{W}$ be a weighted adjacency matrix for an undirected graph $G = (\mathcal{E},\mathcal{V})$ such that $w_{ij}$ is the weight of the edge between vertices $i$ and $j$ in $G$ and $G$ has no multiple edges or self-loops. A simple graph is a graph such that $w_{ij}\in\{0,1\}$ for all $i$ and $j$ and $w_{ij} = 0$ for $i=j$ , where $1$ indicates the existence of an edge. Then $n = |\mathcal{V}|$ is the size of the network and $m$ is the number of undirected edges so that $2m$ is the number of non-zero entries in the symmetric adjacency matrix.
	
	\subsection*{Simulated experiments}
	\subsubsection*{Complex hierarchy models}
	The Weighted Complex Hierarchy (WCH) network takes the existence probabilities of edges in an Erd\" os-R\' enyi random network \cite{Erdos1959} as the base weights of the edges. It then randomly chooses the number of hierarchical levels (between 2 and 5) before randomly assigning nodes to these $h$ levels, based on a geometric cumulative distribution function with a default parameter of 0.6. All of the edges adjacent to nodes in a given level are then provided an addition weight of $(h-1)s$ for some suitably chosen strength parameter, $s$. The weight of an edge in the WCH model is then $\bar{w}_{ij} = w_{ij} + (h(i)-1)s + (h(j)-1)s$, where $w_{ij}$ is distributed uniformly in $[0,1]$ and $h(i)$ is the hierarchical level of node $i$. By fine-tuning the strength parameter, these networks have been shown to mimic the topology of EEG networks formed from the weighted phase-lag index \cite{Smit2016b}. Small differences in parameter determine the strength of the hierarchical structure of the network: s = 0 gives a random network; s= 1 gives a strict 'class-based' topology determined by the node hierarchical levels. Thus, a larger strength parameter provides a network with a more rigid hierarchical structure.
	
	We repeat this methodology for networks with 16, 32, 64 and 128 nodes, spanning a large range of network sizes as used in current research, e.g. see \cite{Tijm2013}.
	
	\subsubsection*{Experimental design}
\begin{figure*}[!htpb]
		\centering
		\includegraphics[trim = 0 150 0 140,clip,scale=0.6]{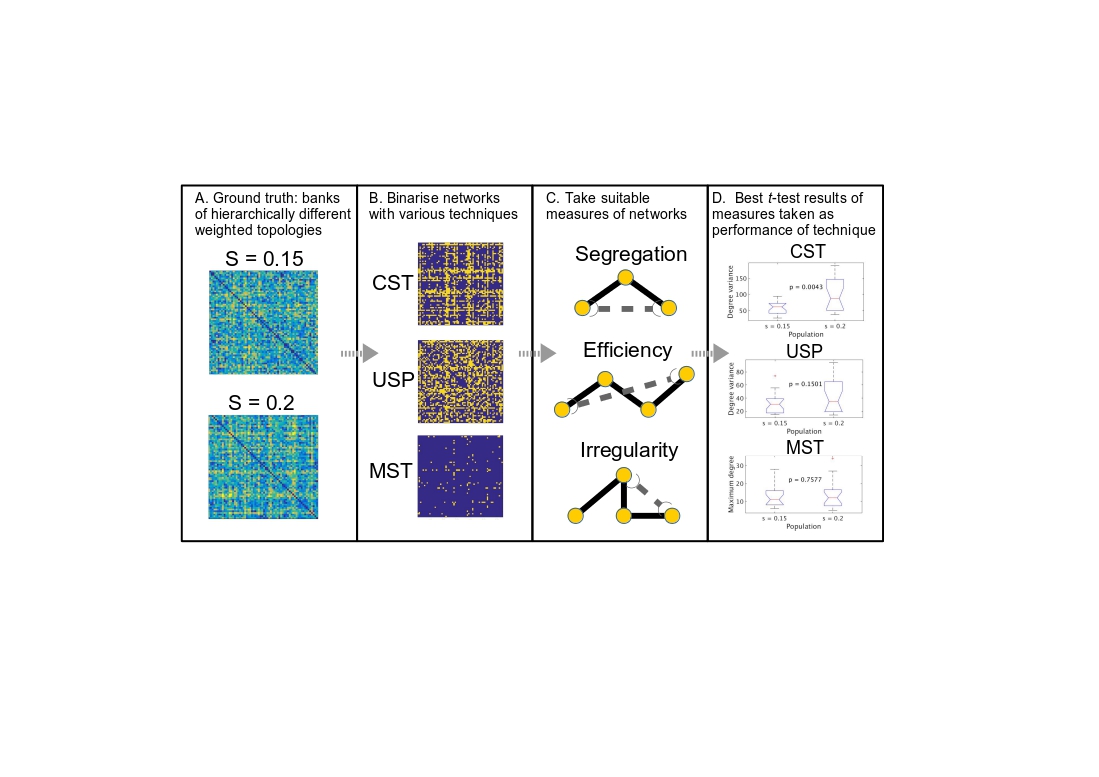}
		\caption{Methodological steps in the evaluation of binarisation techniques for determining ground truth topological differences.}
		\label{methodsExp}
	\end{figure*}

	For the simulations, we follow the procedure as illustrated in Fig \ref{methodsExp}. WCH models are generated as a ground truth to test ability of binarisation techniques to distinguish subtly different populations of size 20. These different populations are generated using realisations of the WCH model with small differences in the strength parameter, $s$. The procedure thus follows that of a typical clinical study, where small populations of contrasted conditions are analysed using network science techniques with statistical tests used to determine significance of the differences between the populations.
	
	\subsubsection*{Random and targeted topological attacks}
	We test the robustness of the given binarisation techniques by subjecting these same simulated networks to random and targeted 'topological attacks' before implementing similar topological comparative analysis as above. Random and targeted attacks were originally formulated by deleting entire nodes from the network \cite{Tan2005}. We implement a weight randomising approach, thus preserving network size which is important when comparing different metrics \cite{Dorm2009}. Further,  this is more relevant to brain networks where the network size is determined a-priori, but rather the information recorded at the nodes are susceptible to attacks. These random topological attacks are implemented by substituting the WCH models weights with corresponding non-zero entries of a sparse, randomly weighted adjacency matrix. We implement this comparison by increasing the density of the sparse matrix, i.e. densities of $0, 0.05, 0.1,..., 0.95,1$. Targeted topological attacks are implemented similarly except the attacks are restricted only to those nodes whose average adjacent weight is over one standard deviation above the mean, relating to those nodes with abnormally strong connectivity. Such strongly weighted nodes are known as `hub' nodes for their importance to the topology of the network.
	
	\subsection*{Network binarisation}	
	\subsubsection*{Minimum spanning tree}
	The MST is a construction based binarisation approach which obtains a tree by selecting the strongest edges of the network such that the network is fully connected and no cycles are present. The algorithm for its construction is well known \cite{Krusk1956} and included in popular toolboxes \cite{Rubi2010}.
	
	\subsubsection*{Union of shortest paths}
	The USP is another construction based approach to unbiased network binarisation \cite{Mei2015}. The shortest path between two nodes in a network is the set of edges with the minimum sum of weights connecting them. This can be constructed using e.g. Dijkstra's \cite{Dijk1959} algorithm to find the shortest paths between each pair of nodes in the network, adding all the edges of those paths to an initially empty binary network. Because connectivity has an inverse relation to distance, the weights of the network must first be relationally inversed in order to construct the shortest paths. This inversion process can take several forms which involves a certain amount of subjective discretion and depends largely on the distribution of the original weights. For our study we choose $\hat{\mathbf{W}} = -\ln(\textbf{W})/\alpha$, where
	\begin{equation}
	\alpha = \text{min}\{\mathbb{N}\} \text{ s.t. } \text{max}_{i,j}( \hat w_{ij}) < 1,
	\end{equation}
	as it has been shown to offer a better spread of metric magnitudes which is important for shortest path problems \cite{Smit2016a}.
	
	\subsubsection*{Cluster-span threshold}
	The CST chooses the binary network at the point where open to closed triples are balanced \cite{Smit2015a}. This balance occurs when $C_{Glob} = 0.5$, which is obvious from the definition. Importantly, the balancing of this topological characteristic necessarily endows the binary network with a trade-off of sparsity to density of edges. We see this since a network is sparse if most triples are open and dense if most triples are closed. It is hypothesised that this balance achieves an informational richness useful for capturing different topologies of EEG functional connectivity \cite{Smit2016b}.
	
	The algorithm for the CST computes the binary networks for each possible number of strongest edges between 15\% to 85\%, rounded to the closest real edge density.	The clustering coefficient is then computed for each of these networks, obtaining a vector $\mathbf{C} = \{C_{15},C_{16},\dots,C_{84},C_{85}\}$,	where $C_{i}$ is the clustering coefficient of the binary network at the $i$th \% density, rounded to the nearest number of edges. Then the network of the CST is the binary network corresponding to $Z = \text{argmin}_{i}(C_{i} - 0.5)$, i.e. the threshold achieving minimum value of the vector $\mathbf{C}$ minus the clustering coefficient value which obtains an equilibrium between triangles and non-triangle triples, 0.5. The values of 15\% and 85\% are chosen as safe values based on experimental evidence \cite{Smit2015a,Smit2016a}. Particularly, below 15\%, real brain networks can have a tendency to fracture into more than one component, thus making calculations of metric values, including the clustering coefficient, inconsistent and unreliable \cite{Stam2007}.
	
	\subsubsection*{Efficiency-cost optimisation threshold}
	The ECO proposes a threshold to keep the strongest $1.5n$ edges (equivalent to a density threshold of 3/(n-1)) which is an approximation based on consistent observations of simulated and real brain networks of the maximum ratio of the combined local and global network efficiencies and density \cite{Fall2017}. It is hypothesised that such a trade-off of network efficiency and sparsity provides networks which are meaningful to the concept of economy in brain function \cite{Bull2012}.
	
	\subsubsection*{Arbitrary proportional thresholds}
	Arbitrary thresholds can be chosen by either choosing a weight above which edges are kept and below which edges are discarded, or by choosing a percentage of strongest weighted edges to keep in the network. The latter choice is more robust and easier to compare between different set-ups and subjects because it keeps the connection density constant and thus is not affected by the values of the weights, which may vary wildly particularly when considering the comparison of different connectivity measures. In order to cover the density ranges of both the sparsity hypothesis and the hierarchical complexity hypothesis, we choose arbitrary thresholds which maintain the strongest 5\%, 10\%, 20\%, 30\%, 40\%, and 50\%  of edges to make sure we cover the relevant array of connection densities whilst reducing redundancy. Note, very sparse densities are already covered by the MST and ECO thresholds.
	
	\subsection*{Network metrics}
	To analyse the simulated and real EEG networks we use a variety of common metrics.
	
	\begin{itemize}
		\item The \textbf{characteristic path length}, $L$, is the average of the shortest path lengths between all pairs of nodes in the graph \cite{Newm2010}.
		\item The \textbf{efficiency}, \textit{Eff} of a weighted network	is the mean of one over the shortest path lengths, thus inversely related to $L$ \cite{Rubi2010}.
		\item The \textbf{diameter}, $D$, of a graph is the largest shortest path length between any two nodes in the graph \cite{Tew2014}.
		\item The \textbf{clustering coefficient}, $C$, is the mean over $i$ of the ratio of triangles to triples centred at node $i$ \cite{Newm2010}.
		\item The \textbf{weighted clustering coefficient}, $C_{\mathbf{W}}$, is a weighted version of the clustering co-efficient for binary networks.
		\item The \textbf{leaf fraction}, \textit{LF}, of a tree is the fraction of nodes in the graph with degree one. Note, every path containing such a node either begins or ends at that node \cite{Tew2014}.
		\item The \textbf{edge density}, $P$, is the ratio of the number of edges in the graph to the total possible number of edges for a graph with the same number of nodes, i.e. $P = 2m/n(n-1)$ \cite{Newm2010}. For the CST, $P$ takes an inversely relational position to $C$ of proportional thresholds. This can be seen by considering two weighted networks whose values of $C$ increase monotonically with increasing $P$ and such that one has higher values of $C$ than the other, which is a working assumption in our case. Then the network with the greater values of $C$ will attain its CST at a lower density, $P$. In a similar vein, $P$ of the USP is inversely related to $L$ of proportional thresholds- the higher the density of the USP, the shorter the average shortest path in the weighted network.
		\item The \textbf{degree variance}, $V$, is the variance of the degrees of the graph which distinguishes the level of scale-freeness present in the graph topology \cite{Smit2016b}.
		\item The \textbf{maximum degree}, \textit{MD}, of a network is just as named- the degree of the node with the most adjacent edges in the network \cite{Tew2014}.
	\end{itemize}
	For each binarisation technique we choose three metrics to analyse the subsequent binary networks. These differ for each technique because of the construction of the network. Particularly, the MST metrics are chosen based on a study of Tewarie et al. \cite{Tew2014}. Similarily, we choose three weighted metrics for analysing the original weighted networks. These choices can be found in Table \ref{metrics}. We try as much as possible to stick to three main categories of metrics for each binarisation technique: segregation (M1 in Table \ref{metrics}), efficiency (M2) and irregularity (M3) \cite{Bull2009,Smit2016b}. This notably deviates for M3 in the weighted case where the mean weight of the network edges, $\mu_{\mathbf{W}}$, is an appropriate and more obvious choice of metric than the variance of those weights.
	
	\begin{table}[h]
		\caption{Grouped Topological Metrics- Three for Each Network Type}
		\label{metrics}
		\centering
		\begin{tabular}{|l+l|l|l|l|l|l|l|}
			\hline
			\textbf{Metric} & \textbf{CST} & \textbf{MST} & \textbf{USP} & \textbf{ECO} & \textbf{Weight} & \textbf{\%T} \\
			\thickhline
			M1 & $P$ & $LF$ & $C$ & $C$ & $C_{\textbf{W}}$ & $C$\\
			\hline
			M2 & $L$ & $D$ & $P$ & $L$ & \textit{Eff} & $L$\\
			\hline
			M3 & $V$ & $MD$ & $V$ & $V$ & $\mu_{\textbf{W}}$ & $V$\\
			\hline
		\end{tabular}
	\end{table}
	
	\subsubsection*{Functional connectivity}
	For the real EEG datasets, FIR bandpass filters were implemented for $\alpha$ (8-13Hz), $\beta$ (13-32Hz) or both bands, as specified in the Materials. The filter order of 70 is used to provide a good trade-off between sharp transitions between the pass and stop bands while keeping the filter order low. The filtered signals were then analysed for pairwise connectivity using a suitable representative of the various phase-based methods \cite{Dauwels2010}, the Phase Lag Index (PLI) \cite{Stam2007b}, to account for the important phase-dependence information seen to underlie electrophysioliogical brain activity \cite{Stam2012}. This avoids the problems of volume conduction found in e.g. correlation or coherence by relying solely on the phase lead/lag information of the signal to negate the redundant correlation effects of synchronous time-series analysis. The PLI between time series $i$ and $j$ is defined as
	\begin{equation}
	PLI_{ij} = |\langle \text{sgn}(\phi_{i}(t)-\phi_{j}(t)) \rangle|,
	\end{equation}
	where the instantaneous phase at time $t$, $\phi_{k}(t)$, is regarded as the angle of the Hilbert transform of the signal $k$ at time $t$. These values were averaged over trials to obtain one connectivity matrix per subject per condition.
	
	\subsection*{Statistical testing}
	In the simulations we undergo 50 simulated trials of two populations of 20 networks for each of a combination of populations. A population of networks is selected from a bank of 1000 WCH networks with given strength parameter. These banks exist for $s = 0$, $0.05$, $0.1$, $0.15$, $0.2$, $0.25$ and $0.3$. The other population in the trial then comes from a WCH bank with strength parameter with 0.05 difference. We undergo such trials for all possible combinations of 0.05 differences. We binarise these networks using each of our binarisation methods. We then compute the three metrics, M1, M2 and M3, for each of these networks (weighted metrics are computed from the original weighted networks). We perform population $t$-tests of these metrics for the WCH binarisations. The assumption of normality of these populations of values are validated with $z$-tests. We choose the best metric of the three to represent the ability of the binarisation method to discern subtle topological differences where the `best' metric is chosen as that which attains the maximum number of significant $p$-values out of the 50 simulated trials which are less than the standard $\alpha = 0.05$ level. If two or more metrics obtain the maximum value, we then choose the one with the lowest mean log of $p$-values. Choosing the log in this instance emphasises the importance of smaller $p$-values for distinguishing differences. The number of differences discovered, taken as a percentage of the total number of trials run, then represents the `accuracy' of the binarisation technique at distinguishing the ground truth, i.e. that the topologies of the populations are different.
	
	\section*{Materials}
	\subsubsection*{Eyes open - eyes closed resting state data}
	The eyes-open, eyes closed 129 node dataset is available online under an Open Database License. We obtained the dataset from the Neurophysilogical Biomarker Toolbox tutorial \cite{NBTdata}. It consists for 16 volunteers and is downsampled to 200Hz. We used the clean dataset which we re-referenced to an average reference before further analysis. The data were filtered in alpha (8-13Hz), according to known effects \cite{Barr2007}, using an order 70 FIR bandpass filter with hamming windows at 0.5Hz resolution. For each subject we take one arbitrarily long 1000 sample epoch (5s) from the 1000-2000th samples, for each subject.
	
	\subsubsection*{Visual short-term memory binding task data}
	We study a 30-channel EEG dataset for 19 healthy young volunteers participating in different VSTM tasks. Full details of the task can be found in \cite{Smit2015b}. Written consent was given by all subjects and the study was approved by the Psychology Research Ethics Committee, University of Edinburgh. The task was to remember objects consisting of either black shapes (Shape) or shapes with associated colours (Binding), presented in either the left or right side of the screen. The sampling rate is 250 Hz and a bandpass of 0.01-40 Hz was used in recording. Based on relevant results \cite{Smit2015a,Smit2015b}, the data were filtered in beta (13-32Hz) using an order 70 FIR bandpass filter with hamming windows at 0.5Hz resolution. The epochs are 1s long and the number of trials is  65.7 $\pm$ 9.27 (mean $\pm$ SD). PLI adjacency matrices are averaged over trials.
	
	\subsubsection*{Alzheimer's disease data}
	The EEG recordings were taken from 12 AD patients and 11 healthy control subjects. The patients -5 men and 7 women; age = 72.8 $\pm$ 8.0 years, mean $\pm$ standard deviation (SD)- were  recruited  from  the  Alzheimer's  Patients'  Relatives  Association of Valladolid (AFAVA). They all fulfilled the criteria for probable AD. EEG activity was recorded at the University Hospital of Valladolid (Spain) after the patients had undergone clinical evaluation including clinical history, neurological and physical examinations, brain scans and a Mini Mental State Examination (MMSE) to assess their cognitive ability \cite{Fol1975}. The ethics committee of the Hospital Clinico Universitario de Valladolid approved the study and control subjects and all caregivers of the patients gave their written informed consent for participation. The 16 channel EEG recordings were made using Profile Study Room 2.3.411 EEG equipment (Oxford Instruments) in accordance with the international 10-20 system. Full details can be found in \cite{EscData}. The data were filtered both in alpha (8-13Hz) and beta (13-32Hz) as in \cite{Smit2016a}, for separate analysis, using an order 70 FIR bandpass filter with hamming windows at 0.5Hz resolution. Recordings were visually inspected by a specialist physician who selected epochs with minimal artefactual activity of 5s (1280 points) from the data for further analysis. The average number of these epochs per electrode per subject was 28.8 $\pm$ 15.5 (mean±SD).
	
	The EEG PLI adjacency matrices used in this study are available at the University of Edinburgh's data depository at http://datashare.is.ed.ac.uk/handle/10283/2783.
	
	\section*{Results}
	\subsection*{Sensitivity to subtle topological differences in synthetic EEG connectivity}
\begin{figure*}[!htpb]
		\centering
		\includegraphics[trim = 20 40 0 0,clip,scale=0.4]{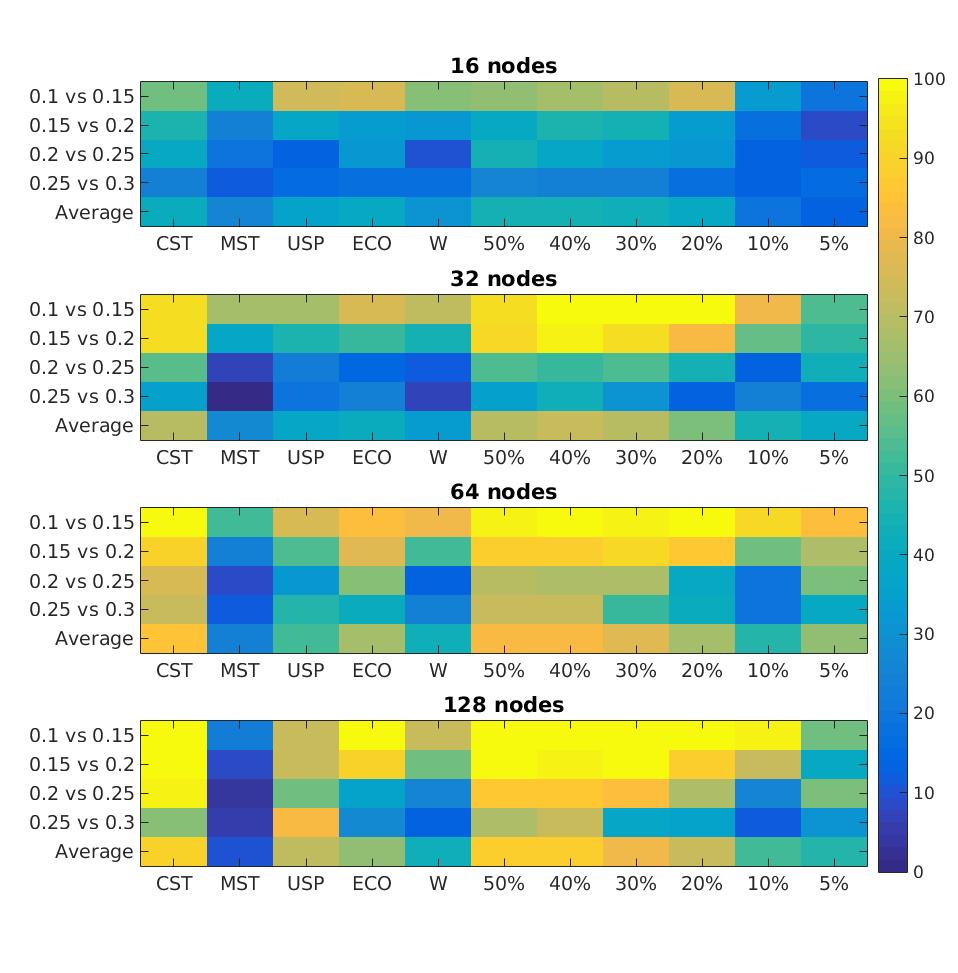}
		\caption{Percentage of topological differences discovered from population $t$-tests between WCH models. The $y$-axis shows $s = a$ vs. $s = b$ for WCH populations with strength parameter, $s$. The $x$-axis shows the binarisation method used where W is the weighted approach and percentages indicate arbitrary density thresholds.}
		\label{WCHPerc}
	\end{figure*}

	Fig \ref{WCHPerc} shows the results for differences discovered between WCH models with differences in strength parameter of 0.05. The grand averages are shown in Table \ref{ModelResults} The CST is shown to outperform other non-arbitrary methods in general. In testing the comparisons of the WCH model with varying strength parameter, $s$, it discovers significant differences at the $\alpha = 5\%$ level 71.3\% of the time over all strength comparisons and network sizes, Table \ref{ModelResults}. On the other hand, the MST discovers just 22.4\% of the differences, the USP discovers 50\% of the differences and the weighted metrics discover just 40.5\% of the differences. Out of these methods, in fact, it discovers the most differences in all but two cases- those being the 0.1 vs 0.15 comparison in the 16 node networks and the 0.25 vs 0.3 node comparison in the 128 node cases, of which the USP is best on both occasions.	

	\begin{table*}[!t]
		\caption{Grand average percentage of topological differences discovered between Weighted Complex Hierarchy models.}
		\label{ModelResults}
		\centering
		\begin{tabular}{|l+l|l|l|l|l|l|l|l|l|l|l|}
			\hline
			\textbf{Method} & \textbf{CST} & \textbf{MST} & \textbf{USP} & \textbf{ECO} & \textbf{Weight} & \textbf{50\% T} & \textbf{40\% T} & \textbf{30\% T} & \textbf{20\% T} & \textbf{10\% T} & \textbf{5\% T}\\
			\hline
			\textbf{Grand Average} & 71.3\% & 22.4\% & 50.0\% & 53.38\% & 40.5\% & 70.4\% & 71.5\% & 67.3\% & 60.0\% & 41.3\% & 41.6\%\\			
			\hline
		\end{tabular}
	\end{table*}
	
	In comparison with arbitrary percentage thresholds the CST appears to perform approximately the same as the 40\% proportional threshold which discerns a slightly higher rate of 71.5\% of differences over all cases. The 50\% threshold also appears to be good at discerning differences here with an overall rate of 70.4\% of differences discovered. These results agree with the hypothesis that complex hierarchical structures are best captured by larger density ranges. It is important to recall that we need non-arbitrary solutions rather than simply to find the best possible threshold for these specific simulations. With this in mind we can see that, compared to the other techniques, the CST outperforms the field in this study.
	
	\subsection*{Robustness to random and targeted topological attacks}
\begin{figure}[!h]
		\centering
		\includegraphics[trim = 0 0 0 0,clip,scale=0.4]{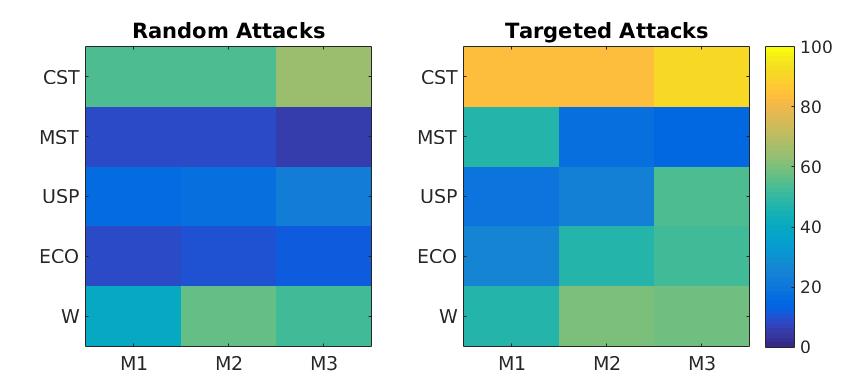}
		\caption{Percentage of topological differences discovered from population $t$-tests via binarisation methods (CST,MST, USP, ECO and weighted network (W)) between WCH models ($s = 0.1$ vs. $s = 0.15$) with random and targeted network attacks. M1, M2 and M3 as in Table \ref{metrics}}
		\label{AttPerc}
	\end{figure}

	The robustness to random and targeted topological attacks is evaluated by comparing the metrics from the attacked WCH models over all non-arbitrary binarisation techniques using population $t$-tests as before. For these analyses we look at the case in Table \ref{ModelResults} with the maximum ratio between the mean and standard deviation of accuracy over binarisation techniques, i.e. the case which maximises the ratio of average performance and comparability of performances. This happens in the 32 node case (6.3538) with differences in strength parameter of $s = 0.1$ and $s = 0.15$. The grand percentage over all sizes of attack for $p$-values below 0.05 for each metric is presented in Fig \ref{AttPerc} for both random and targeted topological attacks. Generally, the binarisation methods as well as the weighted metrics are more robust to targeted attacks than to random attacks. Notably, the CST maintains the highest average accuracy of distinguishing topological differences for all metrics and both random and targeted attacks. This is particularly evident in the targeted attacks. For both kinds of attack, the weighted metrics come in second best while the ECO,  the USP and MST perform relatively poorly.
	
\begin{figure}[!h]
		\centering
		\includegraphics[trim = 0 0 0 0,clip,scale=0.4]{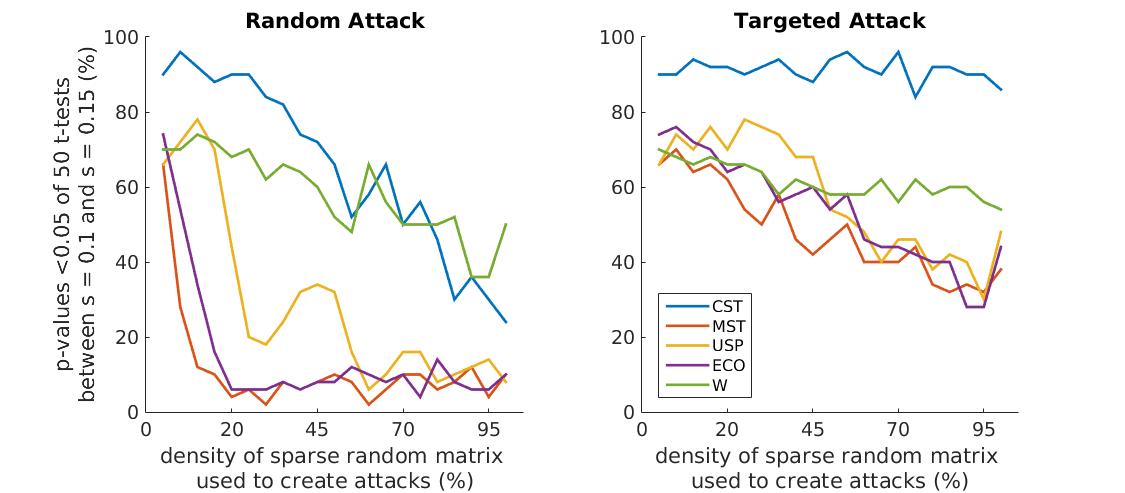}
		\caption{Percentage accuracy of method for distinguishing topological differences between attacked WCH models against the size of those attacks for random topological attacks (left) and targeted topological attacks (right). The values plotted are the maximum from the three metrics, M1, M2 and M3 (as in Table \ref{metrics}), for the corresponding technique as indicated in the legend.}
		\label{AttEach}
	\end{figure}

	The metric achieving highest accuracy for each binarisation technique and for each size of attack is shown in Fig \ref{AttEach} for both random and targeted topological attacks.
		
	Strictly in terms of robustness, as opposed to best accuracy, the weighted networks prove the best, with the least detriment noted by increasing the size of attack in its topological acuity (green). The CST also does well here. For the random topological attacks, even at 50\% of connections attacked, the CST notes an accuracy of 70\% (blue line). The USP (yellow), MST (red) and ECO (purple) networks are not at all robust to random topological attacks in this scenario with immediate drop offs on the implementation of attacks.
	
	For the targeted topological attacks (Fig \ref{AttEach}, right), the CST  network (blue) shows the most resilience with no noticeable depreciation of accuracy. The other methods, in contrast, show a notable decrease in accuracy as more weights are randomised.
	
	\subsection*{Real dataset results}
	We maintain our focus on comparing non-arbitrary methods since arbitrary approaches are inappropriate for neurophysiological studies where one can pick from an order of $n(n-1)/2$ thresholds. 
	
\begin{figure}[!h]
		\centering
		\includegraphics[trim = 50 180 0 180,clip,scale=0.7]{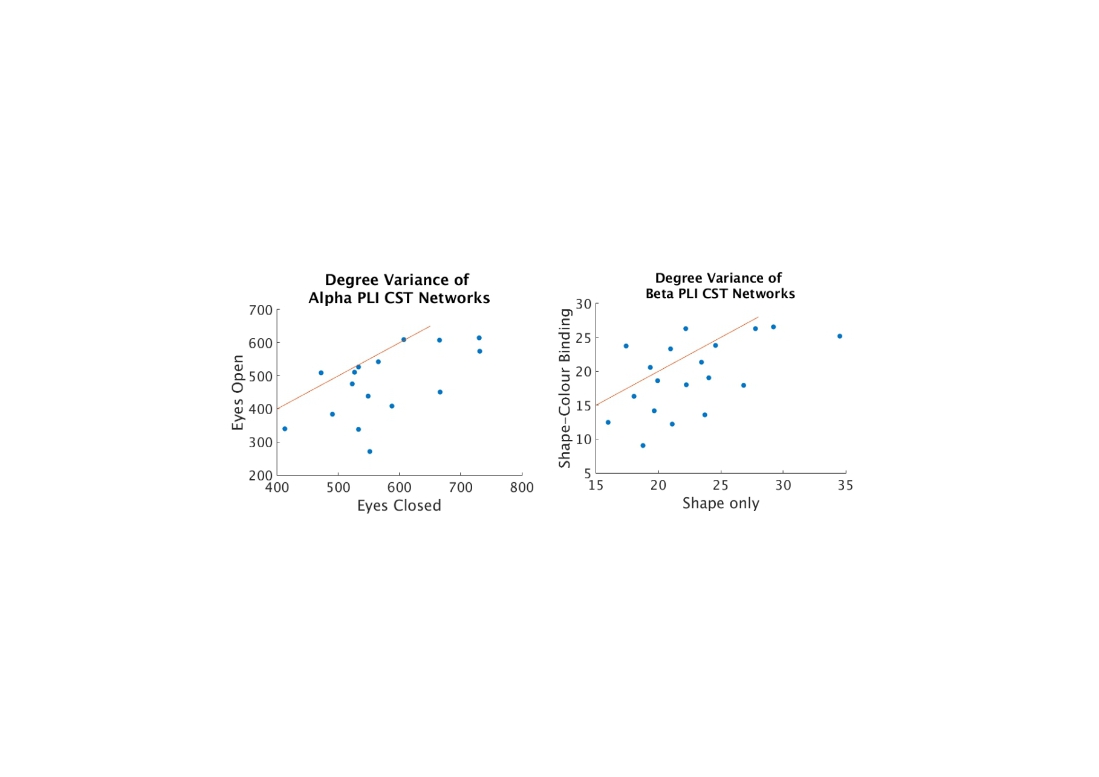}
		\caption{Scatter plots of degree variance for CST networks of Eyes Open vs Eyes closed resting state conditions in $\alpha$, left, and degree variance for CST networks of Shape only vs Shape-colour binding conditions in the Right screen in $\beta$, right.}
		\label{vcst_vstm}
	\end{figure}

	Table \ref{EO_EC} shows the results for distinguishing the difference in $\alpha$ activity well known to exist between eyes-closed and eyes-open conditions \cite{Barr2007}. The CST finds a significant difference in $V$ of eyes open and eyes closed resting state activity indicating that the phase-dependent topology of EEG activity is less scale-free in the eyes-open condition implying greater hub dominance in the eyes-closed condition, see Fig \ref{vcst_vstm}, left. All of the weighted metrics also find significant differences. Neither the MST or USP find any differences between these conditions. Probing further, $\rho(M1,M2)$ being the correlation coefficient of metric values across subjects of metrics M1 and M2 as defined in Table \ref{metrics}, the weighted metrics in this case are all very highly correlated (all $>0.95$ Pearson correlation coefficient, $\rho$, Table \ref{corrs}) within condition. Therefore they cannot be seen to provide any distinct topological information. The corresponding correlations of the CST show a more distinct topological characterisation, see Table \ref{corrs}.
	
	\begin{table}[h]
		\caption{The $p$-values from paired $t$-tests between eyes open (EO) and eyes closed (EC) in $\alpha$-band 129-channel EEG PLI Networks. Underline: Best value for each method. Bold: Significant values. M1, M2 and M3 as in Table \ref{metrics}}
		\label{EO_EC}
		\centering
		\begin{tabular}{|l+l|l|l|l|l|l|} 
			\hline
			\textbf{Metric} & \textbf{CST} & \textbf{MST} & \textbf{USP} & \textbf{ECO} & \textbf{Weight} \\ 
			\thickhline
			M1 & 0.7504 & 0.4178 & 0.5063 & 0.6034 & \textbf{\underline{0.0016}} \\ 
			\hline
			M2 & 0.9319 & 0.4513 & 0.9942 & 0.5281 & \textbf{\underline{0.0034}} \\ 
			\hline
			M3  & \textbf{\underline{0.0006}} & 0.9616 & 0.6577 & 0.6805 & \textbf{0.0016}\\ 
			\hline
		\end{tabular}
	\end{table}	
	
	\begin{table}[h]
		\caption{Pearson correlation coefficients of Metrics in Eyes Open (EO) - Eyes Closed (EC) Dataset for the CST and Weighted Metrics (wgt)}
		\label{corrs}
		\centering
		\begin{tabular}{|l+l|l|l|l|l|}
			\hline
			\textbf{Metric corr.} & \textbf{EC (CST)} & \textbf{EO (CST)} & \textbf{EC (wgt)} & \textbf{EO (wgt)}\\
			\thickhline
			$\rho($M1,M2$)$ & 0.7662 & 0.9692 & 0.9993 & 0.9512 \\
			\hline
			$\rho($M1,M3$)$ & -0.0458 & -0.7301 & 0.9999 & 0.9996 \\
			\hline
			$\rho($M2,M3$)$ & -0.1695 & 0.6677 & 0.9994 & 0.9576 \\
			\hline
		\end{tabular}
	\end{table}
		
	Table \ref{VSTM} shows the results for distinguishing the difference in $\beta$ activity existing between Shape and Binding tasks when tested in the Left and Right sides of the screen separately. A significant difference is found in $V$ of the CST networks in the Right condition. This indicates that the phase-dependent topology of EEG activity is less scale-free in the Binding condition implying greater hub dominance in the Shape condition, see Fig \ref{vcst_vstm}, right. On the other hand, a significant difference ($p = 0.0059$) is found in $C$ for the ECO networks in the Left condition. This indicates sparse density topology of EEG activity is less integrated in the Binding condition.
	
	\begin{table}[h]
		\caption{The $p$-values from paired $t$-tests between Shape only and Shape Colour Binding tasks in $\beta$-band 30-channel EEG PLI networks. Formatting as in Table \ref{EO_EC}}
		\label{VSTM}
		\centering
		\begin{tabular}{|l|l+l|l|l|l|l|l|} 
			\hline
			\textbf{Hemifield} & \textbf{Metric} & \textbf{CST} & \textbf{MST} & \textbf{USP} & \textbf{ECO} & \textbf{Weight} \\
			\thickhline
			& M1 & 0.5128 & 0.7186 & - & \underline{\textbf{0.0059}} & 0.1007 \\ 
			Left & M2 & \underline{0.0898} & 0.1383 & - & 0.1870 & 0.1010 \\ 
			& M3  & 0.8997 & 0.0911 & - & \underline{\textbf{0.0238}} & 0.1010 \\ 
			\hline
			\hline
			& M1 & 0.5877 & \underline{0.1919} & - & 0.5504 & 0.7742 \\ 
			Right & M2 & 0.9196 & 0.5716 & - & \underline{0.5038} & 0.7733 \\ 
			& M3  & \textbf{\underline{0.0088}} & 0.8146 & - & 0.2138 & 0.7733 \\ 
			\hline
		\end{tabular}
	\end{table}
	
	Noticeably, the USP failed to find meaningful network information in this task because, even after transformation, all the weight magnitudes were in a range such that the shortest weighted path between each pair of nodes was the weight of the single edge joining them.
	
\begin{figure}[!h]
		\centering
		\includegraphics[trim = 50 200 0 180,clip,scale=0.7]{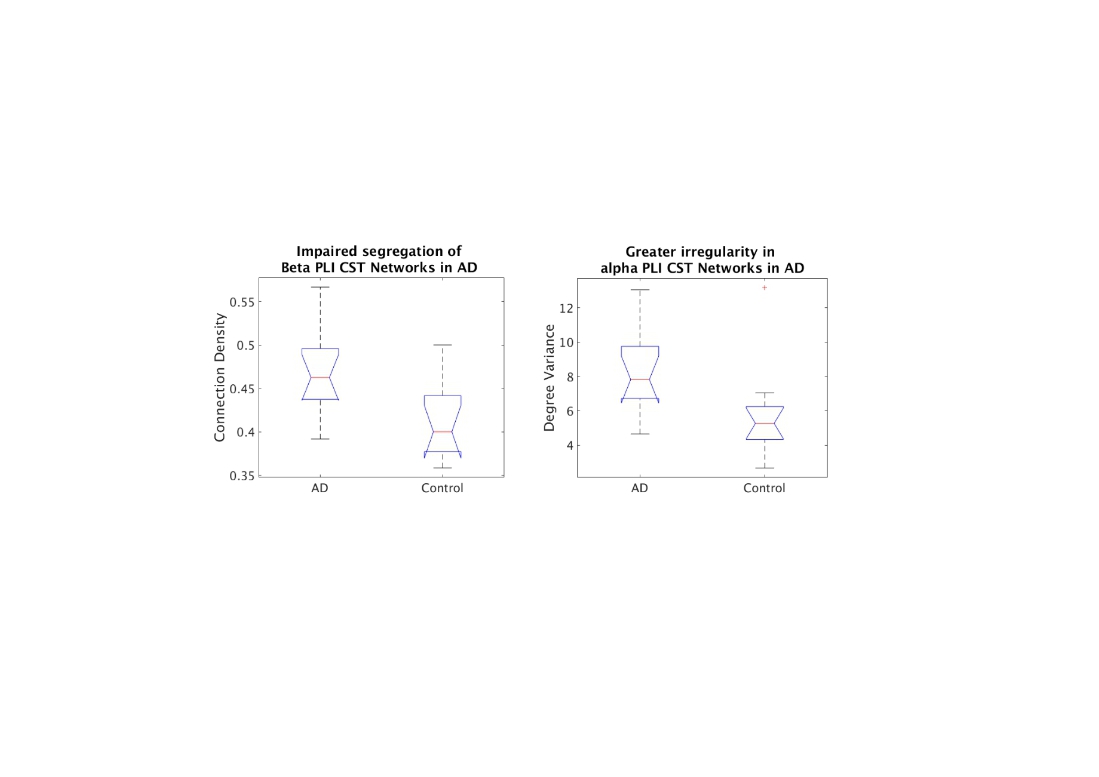}
		\caption{Box plots of connection density, left, and degree variance, right, for CST networks of AD and control in $\beta$ and $\alpha$, respectively.}
		\label{pvcst_ad}
	\end{figure}

	Table \ref{AD} shows the results for distinguishing the difference in both $\alpha$ and $\beta$ activity existing between AD patients and healthy age matched control. For the CST, an effect is noticed in $V$ of $\alpha$ activity (Fig \ref{pvcst_ad}, right) and a larger effect is found in the $P$ of $\beta$ activity (Fig \ref{pvcst_ad}, left). Since $P$ of CST networks is inversely relational to $C$ of arbitrary threshold networks, this tells us that $\beta$ of AD patients is less segregated than control. Contrasting with this, the activity in $\alpha$  suggest a more scale-free network in the alpha band of AD patients than in age matched control.
	
	\begin{table}[h]
		\caption{The $p$-values from population $t$-tests of network measures  AD and control in 16-channel EEG PLI Networks}
		\label{AD}
		\centering
		\begin{tabular}{|l|l+l|l|l|l|l|l|} 
			\hline
			\textbf{Band} & \textbf{Metric} & \textbf{CST} & \textbf{MST} & \textbf{USP} & \textbf{ECO} & \textbf{Weight}\\ 
			\thickhline
			& M1 & \underline{0.0852} & 0.3468 & 0.1167 & 0.7496 & 0.6736 \\ 
			Alpha & M2 & 0.3634 & 0.2630 & 0.1081 & \underline{0.0582} & 0.4189 \\ 
			& M3  & \textbf{\underline{0.0406}} & 0.7324 & 0.0942 & 0.2089 & 0.5570 \\ 
			\hline
			\hline
			& M1 & \textbf{\underline{0.0062}} & 0.4618 & 0.1500 & 0.9775 & 0.7080 \\ 
			Beta & M2 & \underline{0.0529} & 0.6245 & 0.1485 & 0.4946 & 0.4215 \\ 
			& M3  & 0.1782 & 0.5437 & \underline{0.1397} & 0.2575 & 0.5564 \\ 
			\hline
		\end{tabular}
	\end{table}
	
	\subsection*{Density}
	The densities for the datasets used in this study are as in Table \ref{Densities}. For the USP we see both a dependency on network size, where the WCH networks are less dense with increasing size, and on the distribution of weights, where analysis of real datasets becomes implausible since connectivity, averaged over trials, creates a smaller spread and so redundant shortest paths. The MST and ECO are dependent on network size as is obvious from their formulations. The CST binarises the network consistently with a density between 0.3-0.5. From the models we notice that the higher the paramater, $s$, the less dense are the resulting CST networks. Network size appears to have much less effect, which provides evidence to suggest that the CST is dependent on topology, but not on network size.
	
	\begin{table*}[h]
		\caption{Mean and standard deviation (M $\pm$ SD) of densities of CST, USP, MST and ECO networks. WCH\#\# = Weighted Complex Hierarchy model of size \#\#; $s$ = strength parameter of WCH model; PLI\#\# = Phase-Lag Index networks of size \#\#.}
		\centering
		\begin{tabular}{|l+l|l+l|l|l|l|l|}
			\hline
			\textbf{Dataset} & \textbf{Condition} & \textbf{CST} & \textbf{USP} & \textbf{MST} & \textbf{ECO}\\
			\thickhline
			& $s = 0.1$ & $0.49\pm 0.04$ & $0.53 \pm 0.08$ & & \\
			& $s = 0.15$ & $0.46 \pm 0.05$ & $0.59 \pm 0.08$ & & \\
			\textbf{WCH16}  & $s = 0.2$ & $0.44 \pm 0.06$ & $0.62 \pm 0.08$ & 2/16 & 3/15 \\
			& $s = 0.25$ & $0.41 \pm 0.06$ & $0.64 \pm 0.09$ & $= 0.125$ & $= 0.20$\\
			& $s = 0.3$ & $0.40 \pm 0.06$ & $0.64 \pm 0.05$ & & \\
			\hline
			& $s = 0.1$ & $0.48 \pm 0.02$ & $0.45 \pm 0.07$ & & \\
			& $s = 0.15$ & $0.45 \pm 0.03$ & $0.50 \pm 0.06$ & & \\
			\textbf{WCH32}  & $s = 0.2$ & $0.42 \pm 0.04$ & $0.52 \pm 0.07$ & 2/32 & 3/31 \\
			& $s = 0.25$ & $0.39 \pm 0.04$ & $0.53 \pm 0.07$ & $= 0.0625$ & $= 0.0968$ \\
			& $s = 0.3$ & $0.37 \pm 0.04$ & $0.54 \pm 0.08$ & & \\
			\hline
			& $s = 0.1$ & $0.48 \pm 0.01$ & $0.38 \pm 0.05$ & & \\
			& $s = 0.15$ & $0.45 \pm 0.03$ & $0.42 \pm 0.05$ & & \\
			\textbf{WCH64}  & $s = 0.2$ & $0.41 \pm 0.04$ & $0.44 \pm 0.05$ & 2/64 & 3/63 \\
			& $s = 0.25$ & $0.38 \pm 0.04$ & $0.45 \pm 0.05$ & $= 0.0312$ & $= 0.0476$ \\
			& $s = 0.3$ & $0.36 \pm 0.04$ & $0.45 \pm 0.06$ & & \\
			\hline
			& $s = 0.1$ & $0.47 \pm 0.01$ & $0.33 \pm 0.04$ & & \\
			& $s = 0.15$ & $0.45 \pm 0.02$ & $0.35 \pm 0.03$ & & \\
			\textbf{WCH128} & $s = 0.2$ & $0.40 \pm 0.04$ & $0.37 \pm 0.03$ & 2/128 & 3/127 \\
			& $s = 0.25$ & $0.37 \pm 0.04$ & $0.37 \pm 0.04$ & $= 0.0156$ & $= 0.0236$ \\
			& $s = 0.3$ & $0.35 \pm 0.03$ & $0.37 \pm 0.06$ & & \\
			\hline	
			\textbf{PLI16}  & Patient & $0.47 \pm 0.04$ & $1 \pm 0$& 2/16 & 3/15 \\
			& Control & $0.41 \pm 0.05$ & $0.98 \pm 0.05$& $= 0.125$ & $= 0.20$ \\
			\hline
			& Shape Left & $0.41 \pm 0.07$ & $1 \pm 0$ & & \\
			\textbf{PLI30}	& Shape Right & $0.40 \pm 0.06$ & $1 \pm 0$ & 2/30 & 3/29 \\
			& Bind Left & $0.42 \pm 0.06$ & $1 \pm 0$ & $= 0.0667$ & $= 0.1034$\\
			& Bind Right & $0.41 \pm 0.08$ & $1 \pm 0$ & & \\
			\hline
			\textbf{PLI129} & Eyes closed & $0.35 \pm 0.07$ & $0.98 \pm 0.04$ & 2/129 & 3/128\\
			& Eyes open & $0.36 \pm 0.07$ & $0.98 \pm 0.06$ & $= 0.0155$ & $= 0.0234$ \\
			\hline													
		\end{tabular}\label{Densities}
	\end{table*}
	
	\section*{Discussion}\label{discuss}
	From the simulation results of complex hierarchy models we see from proportional thresholds that a larger density range is more effective than sparse models. This agrees with our hypothesis that complex hierarchical models contain a density of information beyond what sparse levels of binarisation can reveal. The fact that the real results for EEG datasets confirm the results in simulations provides further strength to the argument that EEG functional connectivity is highly hierarchically complex and so that sparsity is not always the best working hypothesis for functional brain networks. Other evidence in EEG studies alluding to the benefit of medium density ranges has been documented \cite{Jal2016,Stam2007,Smit2015b}. For a physiological rationale for medium densities we can, for instance, regard function as not only emerging through physical connections as hypothesised in the sparsity hypothesis, but through globally interdependent synchronisations via rapid interregional communications.
	
	Binarised networks generally outperformed weighted approaches in our simulations. Furthermore, weighted network metrics should be used with caution. Particularly, we advise checking their correlations with the mean connection strength.
	
	Other efforts looking to study the role of less strong connections in brain networks have also considered intermediate thresholding by considering networks constructed from connections within different ranges of connectivity strength \cite{Schw2011,Sant2014}. However, studying such topologies is hindered by the fact that the true overlying hierarchical structure becomes hidden. Nodes having more edges in an intermediate level does not, for instance, indicate that that node is a hub, but rather that most of its connections lie within the given range. That is to say that intermediate connections maybe interesting to study, but constructing network topologies from them for analysis is rather obscure. The role of the weakest connections, or, if you will, topological gaps of brain connectivity is also an active area of research \cite{Sant2014}. One can consider, in fact that medium density binarisation does much more to account for such features than sparse binarisations since these gaps become much more defined in higher densities.
		
	As an important addition, the results show that the random topological attacks, rather than targeted topological attacks, are the most effective at deconstructing the topology of our simulations. This perhaps seems counter-intuitive, but can be explained by the fact that only the very top levels of the hierarchy are attacked in the targeted setting, whereas the topology in the remaining levels remains largely intact and, in fact, a new `top level' emerges, maintaining the differences exhibited in the strength parameter, $s$, between the two sets of topology. In fact this agrees with previous functional connectivity studies which detailed the greater resilience of functional brain networks to targeted attacks \cite{Acha2006,Joy2013}. These simulations thus provide the clues as to how the hierarchical structure of functional brain networks play a vital role in this resilience.
	
	The results for $V$ in both the Eyes open vs closed and Shape vs Binding datasets combined can explain that more intensive stimulation (eyes open and Binding) leads to a drop in network efficiency where more localised activity is required for higher functional processing \cite{Bull2009,Stam2014b}. The results for the AD dataset indicate both the increased power in binarisation with the CST compared to other approaches and highlights the importance of binarisation itself for distinguishing dysfunctional AD topology.
	
	AD network studies, over varying platforms, network sizes and density ranges, have been found to show seemingly contrasting results \cite{Tijm2013}. Particularly, Tijm's et al. \cite{Tijm2013} reported that these studies were at different density ranges, and in many cases the density range was simply not recorded. Importantly, no functional studies reported density ranges over 25\%. Nonetheless, we note that our results are in agreement with a 149 node MEG PLI study by Stam et al. \cite{Stam2009}, showing lower clustering in AD than control (density not recorded). This is indicative of a move to a more random topology \cite{Watt1998}.
	
	In terms of network size our simulations suggest that the larger the networks are, the more likely it is that topological differences will be picked up by commonly used metrics. This trend is bucked by the MST for which there is a marked drop off from 32 nodes to 128 nodes. This, however may be explained by the fact that at 32 nodes, the MST makes up $2/n = 6.25\%$ of all possible connections whereas at 64 nodes this percentage is $3.12\%$ and for 128 nodes it is just $1.56\%$ which is in line with the previous discussion that lower network densities can inhibit the ability to find topological differences.
	
	The CST is presented here as a sensitive and powerful binarisation technique for network modelling of EEG phase-based functional connectivity. In simulations it performed to a high standard in all network sizes and topological comparisons as well as in robustness to topological attacks. This was echoed in the results of the real data sets where it was consistently able to identify differences in the presented conditions with not obviously correlated metrics. From the simulation results we can infer a large part of this ability to the density range in which the CST binarises the network. We must note, of course, that all of our real data were from EEG recordings and thus we are cautious of similar comparisons for e.g. fMRI. Indeed, we must acknowledge the limitations and narrow focus of this study for EEG PLI networks. Further, although we conjecture that hierarchical complexity of network topology may be behind the CSTs success, there remain unanswered questions and there is certainly scope for better topological thresholds to be developed based on such hypotheses. We hope this study will stimulate interesting discussions and inspire future research in this direction.
	
	The MST is seen to be robust to fluctuations of the underlying network \cite{Tew2014}. It holds appeal in studies where sparsity is desirable, showing utility in a number of other studies \cite{Lee2010, Schoen2011, Boersma2013, Engels2015}, although it must be noted that in these studies it was not compared with other binarisation methods. In our study, however it appears ineffective and particularly so in larger networks which noticeably corresponds to the MST making up less and less of the connection density as the network grows. This concurs with a recent study where we argued that the robustness to fluctuations also means a poverty of information, supported by evidence from an EEG dataset of cognitive tasks \cite{Smit2015b}.
		
	The USP is the set union of those edges which form the shortest paths between all possible pairs of electrodes. Since, in general, all weights of a functional connectivity network lie between 0 and 1, it is likely that a large percentage of the shortest paths in the network will be constituted of just the single edge joining those nodes. Thus, we can expect very high density networks which only differ in topology by the weakest connections. Indeed, in the original paper \cite{Mei2015} the authors did not implement any transformation of the weights and reported densities above 90\%. To try and counter this unwanted outcome we used a negative exponential transform of the weights before extracting the union of shortest paths, however, in the end it appeared that this was limited in its ability to mitigate the flaws of this method. This was most apparent in the VSTM tasks where it turned out that every shortest path was just the edge between each pair of nodes, redundantly returning complete networks. We believe that further work would need to be done regarding the reliability of the USP in order to make it of use to the neuroscience community.
	
	Although generally outperformed by the CST, we note an agreement with the introductory paper of the ECO threshold \cite{Fall2017} that it generally outperforms the MST. Therefore, we would recommend it over the MST in cases where sparse densities appear more important. It was also able to detect differences in the left hemifield condition of the VSTM dataset. The fact that a mutually exclusive difference was found in the right hemifield condition with the CST suggests the interest in considering how different density ranges may reveal different topological traits of conditions. For example, one may conclude from these results that the backbone of the network is effected by binding in the left hemifield, but that in the right hemifield the binding effect is notable rather in the `fleshed out' regions of the network.
	
	We note that research is also undertaken into statistical methods on expected values of connectivity methods to threshold the network \cite{Fall2014}. However, rather than resolving the arbitrary choice problem, it merely diverts it towards the statistical significance paradigm, where arbitrary standards (e.g. $\alpha = 0.05$) have long been adopted to mitigate an intractable problem. In our case we can rely on graph topological techniques so the problem is not intractable. Moreover, as we have seen, it can be more beneficial to include a large number of edges since it allows for richer information coming from the hierarchical relationships of lower degree nodes. Further problems with a statistical approach relate to difficulties in finding the correct solutions for the numerous new connectivity measures available in a way that is consistent and reliable, biases from the size of available data, and, in the case of data surrogate methods, biases due to network size \cite{Fall2014}. Other researchers look towards the integration of different density ranges, however such an approach will have a tendency towards diluting the potency of potential differences \cite{Gin2011} or falling prey to the multiple comparison problem. We note that, without a ground truth, we rely on the assumption that contrasting conditions provokes a contrast in functional connectivity. Further, although small sample sizes are common in the literature, we fully encourage the move towards larger sample sizes from which more powerful and robust results can be obtained. We find the PLI to be reliable and straightforward to interpret but we recognise that finding appropriate connectivity measures is a much debated topic with many considerations including hypotheses of how brain function takes place; the part and frequencies of the signals that should be used for a given paradigm; whether the measure should provide directedness; and whether the signals should be orthogonalised or relocated to the source space.
	
	\section*{Conclusion}
	The hierarchical topologies of simulated weighted complex hierarchy models and several real datasets of EEG functional connectivity assessed from phase dependencies are found to be well characterised by non-arbitrary binarisations using the CST and arbitrary density binarisations in the range of 40\%-50\%. It is conjectured that this is due to their topologies in this range accounting for a wider range of hierarchical structure, i.e. not just the connectivity in the largest degree nodes. The CST and weighted networks were shown to be robust to random and targeted topological attacks when compared with MST, USP and ECO graphs. In three real datasets constituting varied neuroscientific questions, the medium density range which the CST occupies does indeed appear to be useful with other evidence showing that the ECO is useful in sparse densities. Considering both sparse and larger densities in tandem may prove a more effective way forward than either on their own. We were able to successfully identify different topologies in resting states, in VSTM cognitive tasks and in AD patients compared with control with a notable performance from the CST. This study also validates the WCH model as a sensitive tool for topological comparisons of great relevance to the EEG.
	
	\section*{Acknowledgements}
	We would like the thank Dr. Mario A. Parra (MAP) for providing the VSTM data. This data was supported by Alzheimer's Society, Grant \# AS-R42303 and by MRC grant \# MRC-R42552. We would like to thank Dr. Pedro Espino (Hospital Clinico San Carlos, Madrid, Spain) for his help in the recording and selection of EEG epochs of the AD dataset.
	
	
	%
	%
	%

\end{document}